\documentstyle[12pt]{article}

\def\br{\begin{eqnarray}}
\def\er{\end{eqnarray}}
\def\be{\begin{equation}}
\def\ee{\end{equation}}
\def\d{\delta}
\def\({\left(}
\def\){\right)}

\def\hm{ {-{1\over 2}}  }

\def\l{\lambda}

\def\o{\over}

\begin{document}
\setlength{\baselineskip}{5mm}

\noindent{\large\bf
Supersymmetric Quantum Mechanics, the Variational Method and a
New Shape Invariant Potential
} \footnotemark
\footnotetext{Talk given at the XXIII International Colloquium on Group Theoretical Methods in
Physics, Dubna, August  2000 }
\vspace{4mm}

\noindent{ \underline{Elso Drigo Filho}$^a$ \footnotemark
\footnotetext{e-mail elso@df.ibilce.unesp.br} and \underline{Regina Maria  Ricotta}$^b$  \footnotemark
\footnotetext{e-mail regina@fatecsp.br }}
\vspace{1mm}

\noindent{\small $^a$Instituto de Bioci\^encias, Letras e Ci\^encias Exatas, IBILCE-UNESP,
Rua Cristov\~ao Colombo, 2265 -  15054-000 S\~ao Jos\'e do Rio Preto - SP,
$^b$Faculdade de Tecnologia de S\~ao Paulo, FATEC/SP-CEETPS-UNESP, Pra\c ca  Fernando Prestes, 30
-  01124-060 S\~ao Paulo - SP, Brazil}
\vspace{4mm}

Born 2 decades ago in the study of the SUSY breaking mechanism of higher dimensional
quantum field theories, \cite{Witten}  Supersymmetric Quantum Mechanics (SQM) has so far been
considered as a new field of research, providing not only a supersymmetric interpretation  of the
Schr$\ddot {o}$dinger  equation, but interesting answers in all sorts of non-relativistic quantum
mechanical systems.  Particular points to be mentioned include the better understanding it brought
of  the exactly solvable, \cite{Gedenshtein}-\cite{ Levai},   the  partially solvable,
\cite{Drigo1}, \cite{ Drigo7}, the isospectral, \cite{Dunne} and  the periodic potentials,
\cite{Sukhatme}. Recently the association of the variational method with SQM  formalism has been
introduced to obtain the approximate energy spectra of  non-exactly solvable potentials,
\cite{Gozzi}-\cite{Drigo3}.

Works of reference \cite{Gozzi}  introduce a scheme based in the hierarchy of
Hamiltonians; it permits the evaluation of excited states for one-dimensional systems.
In reference \cite{Drigo3}  a new methodology based in  an {\it ansatz} for the superpotential
which is  related to the trial wave function is proposed. This new  methodology has been
successfully applied to get the spectra of 3-dimensional atomic systems  and it is illustrated
here through systems well fit by the Hulth\'en, the Morse and the  screened Coulomb potentials,
\cite{Drigo3}, \cite{Drigo6}, \cite{Drigo8}. As a byproduct of this investigation a new exactly
solvable potential has been found,  a generalization  of the Hulth\'en potential,  which presents,
in 1-dimension, the property of shape invariance.

Here,  these results in a concise form  and a sketch of the  new
potential  are presented.\\

As we learn from the basis of quantum mechanics,  (see for instance  \cite{Schiff}), the
variational method was conceived to be an approximative method to evaluate the energy spectra of
a Hamiltonian $H$ and, in particular, its ground state. Its central point is the  search for
an optimum  wave-function $\Psi(r)$ that depends of a number of parameters.  This is called the
trial  wave-function.  The approach consists in varying these parameters  in the
expression for the expectation value of the energy
\be
\label{energy}
E = {\int{\Psi^* H \Psi dr}\over  {\int{\mid \Psi \mid^2 dr}}}
\ee
until this expectation value reaches its minimum value.  This value is an upper limit of the
energy level. Even though this method is usually  applied to get the ground
state energy only, it can also be applied to get the energy of the excited states.

Thus most important for the variational method to work is the aquisition of this optimum wave
function.  At this crucial point SQM is used to obtain this function. Based in physical
arguments, an {\it ansatz} for the superpotential is proposed and, through the superalgebra, the
trial wave function is evaluated. By minimizing  the energy expectation value with respect
to a free parameter introduced by the {\it ansatz} the minimum energy is found.

Consider a system described by a given potential $V_1$. The associated  Hamiltonian $H_1$ can be
factorized in terms of  bosonic operators, in $\hbar = c = 1$ units, \cite{Sukumar}-
\cite{Cooper1}.

\be
H_1 =  -{1\o 2}{d^2 \o d r^2} + V_1(r) =  A_1^+A_1^-  + E_0^{(1)}
\ee
where $ E_0^{(1)}$ is the lowest eigenvalue.  Notice that the function $V_1(r)$ includes the
barrier potential term.  The bosonic operators are  defined in terms of the so called
superpotential $W_1(r)$,
\be
A_1^{\pm} =  {1\o \sqrt 2}\left(\mp {d \o dr} + W_1(r) \right) .
\ee
As a consequence of the factorization of the Hamiltonian $H_1$,   the  Riccati equation must be
satisfied,
\be
\label{Riccati}
W_1^2 - W_1'=  2V_1(r) - 2E_0^{(1)}.
\ee
Through the superalgebra, the eigenfunction for the lowest state is related to the superpotential
$W_1$ by
\be
\label{eigenfunction}
\Psi_0^{(1)} (r) = N exp( -\int_0^r W_1(\bar r) d\bar r).
\ee

What is clear is that if the potential is non-exactly solvable, the Hamiltonian is not exactly
factorizable, in other words, there is no  superpotential that satisfies the Riccati equation.
The Hamiltonian can however be factorized by an effective potential.  In this case, the Riccati
equation is exact. What we do is to make an   {\bf ansatz}  for the superpotential  using
physical arguments that approximate the effective potential to the true potential. Having a
superpotential we use the  superalgebra to evaluate the wave function which will definitely
depend on a free parameter, the variational parameter.

We stress that in fact we are dealing with an effective potential $V_{eff}$ that satisfies
Riccati equation, i.e.,
\be
V_{eff}(y) = {\bar W_1^2 - \bar W_1'\o 2}+ E(\bar\mu)
\ee
where $ \bar W_1 = W_1(\bar\mu)$ is the superpotential that satisfies (\ref{Riccati}) for
$\mu = \bar\mu$, the parameter that minimises the energy of eq.(\ref{energy}).\\

\noindent {\bf The Hulth\'en Potential} \\

The Hulth\'en Potential, in atomic units, is given by
\be
\label{Potential}
V_H(r) = - {\d e^{-\d r}\o 1-e^{-\d r}},
\ee
where $\d$ is the screening parameter. This potential has been used in several
branches of Physics, (see \cite{Varshni} and references therein). Its  Hamiltonian is written as
\be
\label{Hulthen}
H = \hm{d^2 \o dr^2} - {\d e^{-\d r}\o 1-e^{-\d r}} + {l(l+1) \o 2 r^2}.
\ee

The associated Schr$\ddot {o}$dinger equation is only solved in closed form for the case $l =
0$, ($s$ waves), \cite{Lam}. Other than this case, the potential barrier term prevent us to solve
the Schr$\ddot {o}$dinger equation and, in SQM superalgebra language, to build the superfamily as
can be done when $l = 0$. This case however serves as a  basis to construct a superpotential
for the $l\not=0$ case, \cite{Drigo3},
\be
\label{W_1H}
W_1(r) = B_1 { e^{-\d r}\o 1 - e^{-\d r}} + C_1
\ee
where
\be
B_1 = - {\d \o 2} (1 + \sqrt {1 + 4l(l+1)}), \;\;\;\;\; C_1 = - {\d \o 2}
{B_1 + 2\o B_1}.
\ee
This in fact defines, through the Riccati equation, an effective
potential whose functional form is
\be
\label{V_1H}
V_{eff}(r) = - {\d e^{-\d r}\o 1-e^{-\d r}} + {l(l+1)\o 2}{{\d}^2 e^{-2\d r}\o
(1-e^{-\d r})^2} + C_1^2.
\ee
We note that, for small values of  $\d$, the second term of (\ref{V_1H})  gives us a
potential barrier term of (\ref{Hulthen}) in first approximation.

For the state $2p$ we use equation (\ref{W_1H}) with $l=1$ and evaluate the wave function,
changing
$\d$ by the variational parameter $\mu$, i.e.,
\be
\label{EH}
\Psi_{\mu} = \Psi_0^{(1)}(r,\mu) = (1 - e^{-\mu r})^{-{B_1 \o \mu}} e^{-C_1 r}.
\ee
The energy  is obtained by minimisation with respect to $\mu$. Thus, the
equation to be $\;\;$ minimised is
\be
\label{energymu}
E_{\mu} = {\int_0^{\infty} \Psi_{\mu}(r) [\hm {d^2 \o dr^2} - {\d e^{-\d r}\o
1-e^{-\d r}} + {l(l+1)\o 2r^2}] \Psi_{\mu}(r) dr
\o \int_0^{\infty} \Psi_{\mu}(r)^2 dr}.
\ee
and the integration is carried out numerically. Our explicit values for the $2p,\; (l=1)$ energy
states for some values of the parameter $\d$ are listed bellow  in Table 1. They are shown
together with direct numerical integration data.
\vskip 1cm

\begin{tabular}{|l|c|c|c|c|c|} \hline
\multicolumn{1}{|l}{State} &
\multicolumn{1}{|c} {Delta} &
\multicolumn{1}{|c} {Variational result}   &
\multicolumn{1}{|c|} {Numerical Integration }  \\  \hline
2p & 0.025 & -0.112760  & -0.1127605   \\ \hline
 & 0.050 & -0.101042  & -0.1010425   \\ \hline
 & 0.075 & -0.089845  & -0.0898478  \\ \hline
 & 0.100 & -0.079170  & -0.0791794  \\ \hline
 & 0.150 & -0.059495   & -0.0594415 \\ \hline
 & 0.200 & -0.041792   & -0.0418860 \\ \hline
\end{tabular}

\vskip 1cm
Table 1. Energy eigenvalues as a function of the screening parameter for the
states $2p$, [eq.(\ref{EH})]. Comparison is made with numerical data of
Ref.\cite{Varshni}.  \\

\noindent {\bf The Morse Potential}\\

The three dimensional Morse oscillator, suitable to describe a diatomic system,  can be
written as,
\be
V_M = D (e^{-2a(r - r_e)} - 2e^{-a(r - r_e)})
\ee
where $D$ is the dissociation energy, $r_e$ is the equilibrium internuclear distance and $a$  is
the range parameter. We rewrite the original Schr$\ddot {o}$dinger equation $H\Psi = E\Psi$ in
terms of a new variable
$y$,
\be
\left(-{d^2 \o d y^2} + {l(l+1) \o y^2} + \l^2( e^{-2(y - y_e)} - 2e^{-(y - y_e)})\right)\Psi(y) =
\epsilon \Psi(y)
\ee
where $y = ar$ and the constants are set like
\be
y_e = ar_e\;,\;\;\;\;\l^2 = {2m D \o a^2 \hbar^2}\;,\;\;\;\;E
= \epsilon {\hbar^2 a^2 \o 2m}
\ee
and the parameter $m $ is the reduced mass of the molecule.\\

For the case $l=0$ the system is again exactly solvable and it   is used to provide information
to the {\it ansatz} to be made to the superpotential for the $l\not=0$ case, for which  an
analytical exact solution cannot be determined.  The superpotential takes the form
\be
\label{W_1M}
W_1(y) = -\l e^{-(y - y_e)} - {(l+1) \o y} + C.
\ee
The argument is that the first term is taken from the one-dimensional resutls, \cite{Drigo4},
the case of $l=0$.  The knowledge of the second term comes from the study of  three-dimensional
potentials, \cite{Drigo5}.  The related effective potential is given by
\be
V_{eff} = -\l e^{-(y - y_e)} + \left( -\l e^{-(y - y_e)} + \mu - {l+1 \o y}\right )^2 -  {l+1 \o
y} + E(\bar\mu)
\ee

The eigenfunction obtained from eq. (\ref{W_1M}) is then
\be
\Psi(y) \propto e^{-\l e^{-(y - y_e)}}\;  y^{l+1} \;  e^{-Cy} .
\ee

Using this expression as a trial wavefunction in the variational method we change the parameter
$C$ by the variational parameter $\mu$, i.e.,
\be
\Psi_{\mu} = \Psi(y, C = \mu) \propto e^{-\l e^{-(y - y_e)}}\;  y^{l+1} \;
e^{-\mu y}.
\ee
The energy  is then obtained by minimisation of the energy expectation value  with respect to $\mu$.
Thus, the equation to be  minimised is
\be
\label{energymu}
E_{\mu} = {\int_0^{\infty} \Psi_{\mu}(y) [- {d^2 \o dy^2} + \l^2( e^{-2(y - y_e)} - 2e^{-(y -
y_e)}) + {l(l+1)\o y^2}] \Psi_{\mu}(y) dy
\o \int_0^{\infty} \Psi_{\mu}(y)^2 dy}.
\ee

We have used this expression to minimize the energy expectation value of
various molecules: $H_2$, $HCl$, $CO$ and $LiH$, \cite{Drigo6}.  The explicit values of the
energy  for $n=0$ and different values of $l$ are shown bellow for the $H_2$ molecule, for
known values of their respective potential  parameters, \cite{Varshni2}: $D$, $a$, $r_e$ and
$m$. \\
\vskip .5cm
\begin{tabular}{|l|c|c|c|c|c|} \hline
\multicolumn{1}{|c} {l} &
\multicolumn{1}{|c} {Variational results }   &
\multicolumn{1}{|c|} {Shifted $1/N$ } &
\multicolumn{1}{|c|} {Modified shifted $1/N$ }  &
\multicolumn{1}{|c|} {Exact Numerical} \\
\multicolumn{1}{|c} {} &
\multicolumn{1}{|c} {} &
\multicolumn{1}{|c} {expansion results } &
\multicolumn{1}{|c} {expansion results } &
\multicolumn{1}{|c|} {}  \\ \hline
 0 & -4.4758  & -4.4749  & -4.4760 & -4.4759 \\ \hline
 5 & -4.2563  & -4.2589  & -4.2592 & -4.2589\\ \hline
10 & -3.7187  & -3.7247  & -3.7252 & -3.7242\\ \hline
 15 & -2.9578  & -2.9663 & -2.9670 & -2.9654\\ \hline
20 & -2.0735   & -2.0839 & -2.0846 & -2.0826\\ \hline
\end{tabular}\\
\vskip .5cm
{\bf Table 2.} Energy eigenvalues (in $eV$) for different values of $l$ for
$H_2$ molecule, with $D = 4.7446 eV$, $a = 1.9426 \AA^{-1}$, $r_e = 0.7416
\AA$ and
$m = 0.50391amu$. Comparison is made with results from  ref. \cite{Morales}.\\

\noindent {\bf The Screened  Coulomb Potential}\\

The screened Coulomb potential is given, in atomic units, by
\be
\label{Coulomb}
V_{SC} = - {e^{-\d r} \o r}
\ee
where $\delta$ is the screened length. The associated radial Schr$\ddot
{o}$dinger equation includes the potential barrier term and it is given by
\be
\label{Hamiltonian}
 \left(-{1\o 2}{d^2 \o d r^2} - {e^{-\d r} \o r} + { l(l + 1) \o
2r^2}\right) \Psi  = E \Psi
\ee
where the unit length is $\hbar^2 /me^2 $ and the energy unit is
$\epsilon_0 = - me^4/\hbar^2$.

In order to determine an effective potential similar to the potential in
the Hamiltonian (\ref{Hamiltonian}), that is the screened Coulomb potential
plus the potential barrier term, the following {\it ansatz} to the superpotential is suggested
\be
\label{superpotential}
W(r) = - (l+1){\d e^{-\d r} \o 1 - e^{-\d r} }+ {1\o  (l+1)} - {\d
\o 2}.
\ee
Substituting it into (\ref{eigenfunction}), one gets
\be
\label{Psi}
\Psi_0 (r) = (1-e^{-\d r})^{l+1} e^{- ({1\o  (l+1)} - {\d\o 2})r}.
\ee
Assuming that the radial trial wave function is given by (\ref{Psi}),
replacing $\d $ by the variational parameter $\mu $, i.e.,
\be
\label{Psimu}
\Psi_{\mu} (r) = (1-e^{-\mu r})^{l+1} e^{ -({1\o  (l+1)} - {\mu \o 2})r},
\ee
the variational energy is given by
\be
\label{energymu}
E_{\mu} = {\int_0^{\infty} \Psi_{\mu}(r) [\hm {d^2 \o dr^2} -
{ e^{-\d r}\o r} + {l(l+1)\o 2r^2}] \Psi_{\mu}(r) dr
\o \int_0^{\infty} \Psi_{\mu}(r)^2 dr}.
\ee
Thus, minimizing this energy with respect to the variational parameter $\mu$ one
obtains the best estimate for the energy of the screened Coulomb potential, \cite{Drigo8}.

As our potential is not exactly solvable, the  superpotential given by eq.(\ref{superpotential})
does not satisfy the Riccati equation (\ref{Riccati}) but it does satisfy it for an effective
potential instead,
$V_{eff}$
\be
V_{eff}(r) = {\bar W_1^2 - \bar W_1' \o 2}+ E(\bar\mu)
\ee
where $ \bar W_1 = W_1(\d=\bar\mu)$ is given by  eq.(\ref{superpotential}) and
$\bar\mu$ is the parameter that minimises the energy expectation value, (\ref{energymu}). It is
given by
\be
\label{effective}
V_{eff}(r) = - {\d e^{-\d r}\o 1-e^{-\d r}} + {l(l+1)\o 2}{{\d}^2  e^{-2\d
r}\o (1-e^{-\d r})^2} + {1\o 2}({1\o l+1} - {\d\o 2})^2  + E(\d),
\ee
where $\d = \bar\mu$ that minimises energy expectation value. One observes
that for small values of
$\d$ the first term is similar to the potential (\ref{Coulomb}) and the last is
approximately the potential barrier term.  This observation allows us to
conclude that the superpotential (\ref{superpotential})  can be used to
analyse the three dimensional screened Coulomb potential variationally through the trial
wavefunction (\ref{Psi}).\\

\begin{tabular}{|c|c|c|c|c|c|c|} \hline
\multicolumn{1}{|c} {} &
\multicolumn{2}{|c} {2p }   &
\multicolumn{2}{|c|} {3d } &
\multicolumn{2}{|c|} {4f }  \\ \hline
{$ \;\;\delta \;\;$}
& {Variational} & {Numerical}   &
{Variational} & {Numerical} &
{Variational} & {Numerical}  \\ \hline
0.001 & -0.2480 & -0.2480 & -0.10910 &-0.10910  & -0.06051 & -0.06052 \\
\hline    0.005 &  - &  - &  - &  - & -0.52930 & -0.05294 \\ \hline
0.010 & -0.2305 & -0.2305  & -0.09212 & -0.09212   &  -0.04419 &  -0.04420
\\ \hline
0.020 & -0.2119 & -0.2119 & -0.07503 & -0.07503 & -0.02897 & -0.02898 \\
\hline
0.025 & -0.2030 & -0.2030  & -0.06714 & -0.06715 & - & - \\ \hline
0.050 & -0.1615 & -0.1615  & -0.03374 & -0.03383 & - &  - \\ \hline
0.100 & -0.09289 & -0.09307   &  - &  - &   - &  - \\ \hline
\end{tabular}\\
\vskip 1cm
{\bf Table 3.} Energy eigenvalues as function of the  screening
parameters $\d$ for $2p$ ($l=1$), $3d $ ($l=2$) and $4f$ ($l=3$) states, in
rydberg units of
energy. Comparison is made with results from  references \cite{Rogers}-\cite{Greene}.\\

{\bf Comments}\\
We have proposed trial wavefunctions to be used in the variational calculation in order to
determine the energy eigenvalues of some atomic systems described by the Hulth\'en,  the Morse
and the screened Coulomb potentials.  These functions were induced from the formalism of SQM. Using physical arguments it is possible to make an {\it
ansatz} in the superpotential which satisfies the Riccati equation by an effective potential.
The superalgebra enables us to take this superpotential and to evaluate the trial wavefunctions
that contains the variational parameter, the parameter that minimises the energy expectation
value of the energy. The results are in very good agreement with the ones available in the
literature.

Thus this new methodology that  associates the variational method with SQM is a simple and good
alternative procedure that enables the evaluation of the energy spectra with reasonable
accuracy.   In particular, when applying the approach to the Hulth\'en potential it was found
that its effective potential was linked to a new exactly solvable potential, that presents, in
one-dimension the property of shape invariance. At that time the focus of our attention was not on
the properties of this new potential, whose main results are presented bellow. \\

{\bf A new shape-invariant potential}\\
As mentioned above, in \cite{Drigo3} when dealing with the evaluation of  the trial wave
function, the following one-dimensional potential, a generalization of the Hulth\'en potential,
was found,
\be
V(x) = A{ e^{-2\d x}\o (1-e^{-\d x})^2} - B{\d e^{-\d x}\o 1-e^{-\d x}}.
\ee
Notice that when setting $A=0$ and $B=1$ the original Hulth\'en potential is recovered, eq.
(\ref{Potential}). For convenience we redefine the constants A and B such that $V(x)$ becomes
\be
\label{V1}
V_1(x) = a_1(a_1-\d){ e^{-2\d x}\o 2(1-e^{-\d x})^2} - a_1(2b_1 + \d){e^{-\d x}\o 2(1-e^{-\d
x})} .
\ee

From the basis of SQM the following superpotential
\be
W_1 = - a_1 { e^{-\d x}\o 1-e^{-\d x} }+ b_1
\ee
factorizes the related Hamiltonian
\be
H_1 =  -{1\o 2}{d^2 \o d x^2} + V_1(x) =  A_1^+A_1^-  + E_0^{(1)}
\ee
and satifies the Riccati equation associated \\
\br
W_1^2(r) - W_1^{'}(x) & = & 2 V_1 - 2 E_0^{(1)}    \nonumber \\
& = & a_1(a_1 - \d) {e^{-2\d x}\o (1-e^{-\d x})^2} -  a_1 (2b_1 + \d)  {e^{-\d x}\o (1-e^{-\d
x})} + b_1^2 \nonumber
\er
where the lowest energy-eigenvalue is given by
\be
E_0^{(1)} = - {b_1^2 \o 2}
\ee
From the superalgebra,  the associated  eigenfunction,  is given by
\be
\Psi_0^{(1)} = (1 - e^{-\d x})^{{a_1 \o \d}} e^{-b_1 x}.
\ee
The condition we must  impose on  the above wave-functions is to
vanish at infinity and at the origin, i.e.,
\be
{a_1 \o \d} > 0\;,\;\;\;\;\; b_1 > 0 \;\;\;\;\;\rightarrow \;\;\;\;\;a_1,
b_1, \d > 0
\ee
This process of factorization can be repeated and the whole hierarchy of this potential can be
evaluated. The general form for the superpotential  is written as
\be
W_{n+1}= - a_{n+1} { e^{-\d x}\o 1-e^{-\d x} }+ b_{n+1}
\ee
which are related to the potentials
\be
\label{NewV}
V_{n+1} = a_{n+1}(a_{n+1} -\d) { e^{-2\d x}\o 2(1-e^{-\d x})^2} - a_{n+1}(2b_{n+1} +
\d){e^{-\d x}\o 2(1-e^{-\d x})}
\ee
with lowest levels given by
\be
E_0^{({n+1})} = - {b_{n+1}^2 \o 2}
\ee
and with the constants being given by
\be
a_{n+1} = a_1 + n\d
\ee
and
\be
b_{n+1} =  {1 \o 2a_{n+1}} (a_1 (2b_1 + \d) - 2\d (na_1 + {n(n-1)\d \o
2})) - {\d \o 2} .
\ee
For the particular case where {\bf $n=0$} and fixing the constants to
\be
a_1 = \d \;\;,\;\;\;\;\;2b_1 +\d = 2
\ee
the original Hulth\'en potential is recovered,
\be
V_1(x) =  -\d {e^{-\d x}\o 1-e^{-\d x}}
\ee
which can also be factorized, as already known, \cite{Drigo3}. Its whole hierarchy of
potentials is given by
\be
\label{Vn}
V_n = {n(n-1) {\d}^2 e^{-2\d x}\o 2 (1-e^{-\d x})^2  } -  {\d(2 + n(1-n)
\d)e^{-\d x} \o  2 (1-e^{-\d x})}
\ee
with lowest energy levels
\be
E_0^{(n)} = - {1 \o 2}({1\o n} -  {n \d \o 2})^2
\ee
and related superpotentials
\be
W_n(x) = - a_{n} { e^{-\d x}\o 1-e^{-\d x} }+ b_{n}
\ee
where the lowest states are
\be
\Psi^{(n)}_0 = (1 - e^{-\d x})^{a_n/\d} e^{-b_n x}
\ee
with constants given by
\be
a_n/\d = n \,\,,\,\,\,\, b_n = {1 \o n} - {n \d \o 2}.
\ee
At this point we remark that the Hulth\'en potential is non-shape-invariant, \cite{Gedenshtein}.
This is clear by observing the $n=1$ and $n=2$ cases of equation (\ref{Vn}), whereas the new
potential clearly is shape invariant, since all the potentials preserve the shape in the
hierarchy eq. (\ref{NewV}). More details shall soon be found in \cite{Drigo9}.

\end{document}